\documentstyle[12pt]{article}

\newcommand{\be}{\begin{equation}}
\newcommand{\ee}{\end{equation}}
\newcommand{\ba}{\begin{eqnarray}}
\newcommand{\ea}{\end{eqnarray}}
\newcommand{\al}{\alpha}

\newcommand{\msc}[1]{\mbox{\scriptsize #1}}
\newcommand{\th}{{\theta}}
\newcommand{\tchi}{\tilde{\chi}}
\newcommand{\df}{\stackrel{\rm def}{=}}
\newcommand{\dsp}{\displaystyle}
\newcommand{\ep}{\epsilon}
\newcommand{\mod}{\mbox{mod}}
\newcommand{\Th}[2]{\Theta_{#1,#2}}
\newcommand{\tK}{\tilde{K}}
\newcommand{\cN}{{\cal N}}

\makeatletter
\@addtoreset{equation}{section}
\def\theequation{\thesection.\arabic{equation}}
\makeatother

\begin{document}
\begin{titlepage}

 {\baselineskip=14pt
 \rightline{
 \vbox{\hbox{hep-th/0108091}
       \hbox{UT-961}
       }}}

\vskip1.5cm

\begin{center}
{\bf \large CFT Description of String Theory Compactified on \\

\bigskip

Non-compact Manifolds with $G_2$ Holonomy}
\end{center}

\bigskip

\vskip2cm

\begin{center}
Tohru Eguchi and Yuji Sugawara \\

\bigskip

Department of Physics, University of Tokyo\\

\bigskip

Tokyo, Japan 113-0033
\end{center}

\vskip1.5cm
\begin{center}
{\it Dedicated to the memory of Sung-Kil Yang} 
\end{center}

\vskip2.5cm

\begin{abstract}
\vskip0.5cm

We construct modular invariant partition functions for strings
propagating on non-compact manifolds of $G_2$ holonomy.
Our amplitudes involve a pair of ${\cal N}=1$ minimal models ${\cal M}_m$,
${\cal M}_{m+2}$ ($m=3,4,\cdots$) and are identified as describing 
strings on manifolds of $G_2$ holonomy associated with $A_{m-2}$
type singularity.
It turns out that due to theta function identities our amplitudes may be 
cast into a form which contain tricritical Ising model for any $m$.
This is in accord with the results of Shatashvili and Vafa. 
We also construct a candidate 
partition function for string compactified on a non-compact Spin(7) manifold.
\end{abstract}

\end{titlepage}


\section{Introduction}

~

\vskip-0.5cm

Recently 7-dimensional manifolds with $G_2$ holonomy are receiving a lot of 
attentions.
These manifolds provide ${\cal N}=1$ 4-dimensional compactifications of 
M-theory
which are of fundamental physical interest.
They also play an interesting role in a novel duality involving gauge and 
gravitational fields.
In the duality conjectured by Vafa \cite{V} Type IIA theory compactified
on deformed conifold with D6-branes and Type IIA theory on resolved conifold
with RR flux are related. This duality has been explained by lifting the
Type IIA configurations to the M-theory backgrounds on three different
manifolds of $G_2$ holonomy which are smoothly connected 
to each other due to quantum effects \cite{Ach,AMV,AW}.
This new type of duality has been further discussed in \cite{SiV}-\cite{DOT}.
In the non-compact cases explicit Ricci flat metrics 
on manifolds of exceptional holonomy ($G_2$ and Spin(7)) 
have been known for some time \cite{BS,GPP}. 
Recently new metrics have been
discussed by various authors \cite{CGLP,BGGG,CGLLT} while the existence 
theorems of metrics in the compact cases are given in \cite{Joyce}.

In the case of string theory compactified on manifolds of exceptional
holonomy we expect a more
detailed description of the dynamics by making use of world-sheet techniques. 
World-sheet description of string theory on these manifolds 
was discussed by 
Shatashvili and Vafa \cite{SV}. In particular they argued that CFT
description of manifolds with $G_2$ holonomy contains the
tricritical Ising model while that of 8-dimensional 
manifolds with Spin(7) holonomy involves the Ising model. 
We recall that the tricritical Ising model with central charge $c=7/10$ is 
the first one in the ${\cal N}=1$ Virasoro (unitary) minimal series 
(or the 2nd in the Virasoro ${\cal N}=0$ minimal series). 
Related results of CFT on manifolds
with special holonomy are given in \cite{OF,Gep}.

In this article we construct modular invariant partition functions
of strings compactified on non-compact manifolds with exceptional
holonomies. We combine ${\cal N}=1$ Liouville field and ${\cal N}=1$ 
minimal models
so that the states in NS and Ramond sectors cancel each other and the theory
possesses space-time supersymmetry. We obtain a series of partition
functions for strings on $G_2$ holonomy manifolds associated with  
$A_n$-type singularities. We also obtain a partition function on a manifold
with Spin(7) holonomy.

In the case of manifolds with $G_2$ holonomy 
our amplitudes involve a pair of ${\cal N}=1$ minimal models
${\cal M}^{\cN=1}_m$, ${\cal M}^{\cN=1}_{m+2}$ for $m=3,4,5,\cdots$.
It turns out that due to theta function identities our amplitudes may be
rewritten into a form which contain tricritical Ising model together with
some ${\cal N}=0$ conformal system for any value of $m$.
Thus our construction is consistent with the characterization of 
Shatashvili and Vafa \cite{SV} for manifolds with $G_2$ holonomy.
We can explicitly check the existence of the (unique) space-time supersymmetry
operator in our partition functions. 
In the case of a non-compact manifold with Spin(7) holonomy we have
a contribution of the superpartner of the Liouville field which is identified
as the Ising model in the description of \cite{SV} of manifolds with
Spin(7) holonomy.

\section{Manifolds with $G_2$ Holonomy}

~
\vskip-0.5cm

We first consider the type II string theories compactified 
on the 7-dimensional manifolds with $G_2$ holonomy.
Our construction of partition function is analogous to 
that for singular Calabi-Yau manifolds \cite{ES,Miz} which makes 
use of the ${\cal N}=2$ Liouville theory. Let us start with recalling briefly 
these constructions.
Results of \cite{ES} may be reproduced starting from an identity
\be
\chi^{(k)}_{\ell}(\tau)\left({\left(\theta_3(\tau)\over \eta(\tau)\right)}^4
-{\left(\theta_4(\tau)\over \eta(\tau)\right)}^4
-{\left(\theta_2(\tau)\over \eta(\tau)\right)}^4\right)=0,
\hskip3mm \ell=0,1,2\cdots,k .
\ee 
Here $\chi^{(k)}_{\ell}(\tau)$ denotes the character of the level-$k$ $SU(2)$
WZW model of spin $\ell/2$ representation. In the above we have simply 
multiplied 
$\chi^{(k)}_{\ell}(\tau)$ to the combination of theta functions which vanish
due to the well-known Jacobi's identity. The factor
$\chi^{(k)}_{\ell}(\tau)$ 
is identified as describing the ALE space of $A_{k+1}$ type.
In fact if we extract a power of Jacobi theta function
$\theta_i (\hskip1mm i=1,2,3)$ out of 
$\theta_i^4$ and multiply it against $\chi^k_{\ell}$, we find
\be
\sum_m {\Theta_{m,k+2}\over \eta}
\left(\left({\theta_3\over \eta}\right)^{\,3}ch^{NS}_{\ell,m}
-\left({\theta_4\over \eta}\right)^{\,3}
\tilde{ch}^{NS}_{\ell,m}-\left({\theta_2\over \eta}\right)^{\,3}ch^R_{\ell,m}
\right)
=0 ,
\label{singcy}
\ee
where $ch^*_{\ell,m}$ denote the characters of ${\cal N}=2$ minimal model of
level $k$ and we have used the product formula of theta functions.
$\Theta_{m,k+2}$ denotes the standard theta function at level $k+2$.
Eq.(\ref{singcy}) is identified as describing the 6-dimensional string theory 
compactified on ALE space of the $A_{k+1}$ type. Fermions in the four 
transverse 
directions in the Minkowski space ${\bf R}^6$ 
together with the two fermions of ${\cal N}=2$ 
Liouville sector constitute the factor $(\theta_i/\eta)^{\,3}$ 
in (\ref{singcy}). $U(1)$ theta function $\Theta_{m,k+2}$ 
stands for the contribution of the momentum sum of the compact bosonic field 
$Y$ in the ${\cal N}=2$ Liouville system \cite{OV,ES}.

If one further extracts $\theta_i$ out of 
$\theta_i^{\,3}$, multiplies it against $\Theta_{m,k+2}$ 
and uses the product formula for theta functions,
one obtains the conformal blocks of strings compactified on Calabi-Yau 3-folds 
with the A-D-E singularities \cite{ES}.
Construction for the Calabi-Yau 4-fold is similar.

Crucial ingredient in these constructions is the relationship 
between the character of $SU(2)$ WZW model and the ${\cal N}=2$ minimal theory
\be
\chi^{(k)}_{\ell}\times \theta_i \Longleftrightarrow ch^*_{\ell,m}\times 
\Theta_{m,k+2} .
\label{ns5}
\ee
(\ref{ns5}) lies behind the duality between the geometry of NS5-branes and ALE
space \cite{OV}.

Now, let us turn to the construction of partition functions 
for the case of $G_2$ holonomy. In this case 
we have to use the ${\cal N}=1$ world-sheet SUSY
rather than ${\cal N}=2$ SUSY. 
In the following we introduce an ${\cal N}=1$ Liouville system 
consisting of a scalar field $\phi$ 
coupled to the background charge and a free Majorana fermion 
field $\psi$ in order to describe non-compact space-time.
Thus the total system consists of 
a CFT describing the geometry of the $G_2$ manifold, 
the ${\cal N}=1$ Liouville theory and 
an additional free boson and fermion associated with the transverse 
direction of the Minkowski space ${\bf R}^3$.

We first recall that the $SU(2)\,(SO(3))$ current
algebra at level $2$ 
may be realized by 3 Majorana fermions. Characters of the level 2 $SU(2)$ 
affine Lie algebra of spin $\ell/2$ are in fact given by
\ba
&&\hskip-10mm \chi_{\ell=0}^{(2)}={1 \over 2} q^{-1/16}\,
\left(\prod^{\infty}_{n=1}(1+q^{n-1/2})^3+
\prod^{\infty}_{n=1}(1-q^{n-1/2})^3\right) = \frac{1}{2}\left\{
\left(\frac{\th_3}{\eta}\right)^{3/2}+\left(\frac{\th_4}{\eta}\right)^{3/2}
\right\}, \\
&&\hskip-10mm\chi_{\ell=2}^{(2)}={1 \over 2}q^{-1/16}\,
\left(\prod^{\infty}_{n=1}(1+q^{n-1/2})^3
-\prod^{\infty}_{n=1}(1-q^{n-1/2})^3\right)=\frac{1}{2}\left\{
\left(\frac{\th_3}{\eta}\right)^{3/2}-\left(\frac{\th_4}{\eta}\right)^{3/2}
\right\}, \\
&&\hskip-10mm\chi_{\ell=1}^{(2)}= 2q^{1/8}\,\prod^{\infty}_{n=1}(1+q^{n})^3
=\frac{1}{\sqrt{2}}\left(\frac{\th_2}{\eta}\right)^{3/2}.
\ea
Thus roughly the $3/2$ powers of the Jacobi theta functions give the
level=2 characters and the
factors $\theta_i^4$ in (\ref{singcy}) may be rewritten as
\be
\theta_i^4=\theta_i\times \chi^{(2)}_{\ell}\times 
\chi^{(2)}_{\ell'} .
\ee
We then recall the standard coset construction of ${\cal N}=1$ Virasoro
minimal series given by
\be
{\cal M}^{\cN=1}_m: \hskip3mm {SU(2)_k\times SU(2)_2\over SU(2)_{k+2}},
\hskip3mm m=k+2.
\ee
At the level of characters one has
\begin{eqnarray}
\left(\frac{\th_3}{\eta}\right)^{3/2}\chi^{(m-2)}_{r-1}
&=& \sum^{m+1}_{\stackrel{s=1}{s-r\equiv 0 ~(\msc{mod} ~2)}}\,
\chi^{(m)\,NS}_{r,s}\,\chi^{(m)}_{s-1} ,  \label{branching N1-1}\\
\left(\frac{\th_4}{\eta}\right)^{3/2}\chi^{(m-2)}_{r-1}
&=& \sum^{m+1}_{\stackrel{s=1}{s-r\equiv 0 ~(\msc{mod} ~2)}}\,
\tchi^{(m)\,NS}_{r,s}\,\chi^{(m)}_{s-1} ,  \label{branching N1-2}\\
\frac{1}{\sqrt{2}}\left(\frac{\th_2}{\eta}\right)^{3/2}\chi^{(m-2)}_{r-1}
&=& \sum^{m+1}_{\stackrel{s=1}{s-r\equiv 1 ~(\msc{mod} ~2)}}\,
\chi^{(m)\,R}_{r,s}\,\chi^{(m)}_{s-1} \label{branching N1-3} .
\end{eqnarray}
$\chi^{(m)\,*}_{r,s}$ denotes the characters of ${\cal N}=1$
minimal model of central charge $c=3/2(1-8/m(m+2))$. 
See appendix for the explicit expressions of ${\cal N}=1$ characters
(\ref{character NS-1}), (\ref{character NS-2}), (\ref{character R}). 

Making use of 
(\ref{branching N1-1}),(\ref{branching N1-2}),(\ref{branching N1-3}) twice we 
can now rewrite (\ref{singcy}) as
\begin{equation}
0\equiv  \chi^{(m-2)}_{r-1}(\tau) 
\times \frac{1}{\eta^4}(\th_3^4-\th_4^4-\th_2^4)
=2 \sum_{s=1}^{m+3} F^{(m)}_{rs}(\tau)\chi^{(m+2)}_{s-1}(\tau) .
\label{identity G2}
\end{equation}
Here the conformal blocks $F_{r,s}^{(m)}$ are defined as
\begin{eqnarray}
&&F^{(m)}_{rs}(\tau)\equiv \sum_{p=1}^{m+1}\,
\left\{{1 \over 2}\frac{\th_3}{\eta}\chi^{(m)\,NS}_{r,p}\chi^{(m+2)\,NS}_{p,s}
-{1\over2}\frac{\th_4}{\eta}\tchi^{(m)\,NS}_{r,p}\tchi^{(m+2)\,NS}_{p,s}
\right\}  \label{cba G2}\nonumber\\
&&\hskip40mm - \sum_{p=1}^{m+1}\,
\frac{\th_2}{\eta}\chi^{(m)\,R}_{r,p}\chi^{(m+2)\,R}_{p,s} ~,
\label{cb G2} \\
&&F^{(m)}_{m-r\,m+4-s}(\tau)=F^{(m)}_{rs}(\tau)~,
\end{eqnarray} 
where $r,s$ run over the ranges $1\leq r\leq m-1$, 
$1\leq s\leq m+3$, and $r+s\equiv 0 ~ (\mbox{mod~2})$. Sum on $p$ runs over
$r-p\equiv 0, s-p\equiv 0 ~ (\mbox{mod~2})$ in NS sector
while $r-p\equiv 1, s-p\equiv 1 ~ (\mbox{mod~2})$ in R sector.

Because of the identity (\ref{identity G2}) the branching
functions $F^{(m)}_{r,s}$ 
are expected to vanish for all values of $m,r,s$,
\be
F^{(m)}_{r,s}=0, \hskip5mm \mbox{all } m,r,s.
\ee 
We have explicitly verified 
this by Maple for smaller values of $m$. 

Thus we have constructed the conformal blocks $F^{(m)}_{r,s}$
for the candidate partition 
function for string theory on $G_2$ holonomy manifolds. Blocks are made of
a pair of ${\cal N}=1$ minimal models ${\cal M}^{\cN=1}_m,
{\cal M}^{\cN=1}_{m+2}$
with central charges 
$c(m)=3/2(1-8/m(m+2)),
c(m+2)=3/2(1-8/(m+2)(m+4))$ and a single 
power of $\theta_i$ being
interpreted as the contribution of a transverse fermion in Minkowski 3-space 
and the fermion $\psi$ 
of the ${\cal N}=1$ Liouville sector. Total central charge is
given by
\be
(1+{1\over 2})+(1+3Q^2+{1 \over 2})+c(m)+c(m+2)=12.
\ee
Here $Q$ is the Liouville background charge which is adjusted to the value
\begin{equation}
Q=2\sqrt{\frac{1}{2}+\frac{2}{m(m+4)}}.
\end{equation}

As is obvious from the construction, blocks $F^{(m)}_{r,s}$ have 
a good modular properties
\begin{equation}
F^{(m)}_{rs}(-\frac{1}{\tau}) = 
\sum_{r'=1}^{m-1}\sum_{\stackrel{s'=1}{|s'+r'|\equiv 0~(\msc{mod}~2)}}^{m+3}\,
S^{(m-2)}_{rr'}S^{(m+2)}_{ss'}\, F^{(m)}_{r's'}(\tau)~,
\label{S Frs}
\end{equation}
\begin{equation}
F^{(m)}_{rs}(\tau+1) = \exp\left\{2\pi i\left(
\frac{1}{3}+\frac{r^2}{4m}-\frac{s^2}{4(m+4)} \right)
\right\} \, F^{(m)}_{rs}(\tau)~.
\label{T Frs}
\end{equation}
Here $\dsp S^{(k)}_{rr'}\equiv 
\sqrt{\frac{2}{k+2}}\sin\left(\frac{\pi rr'}{k+2}\right)$ denotes
the modular matrix of $SU(2)_k$. The following combination of conformal
blocks gives a modular invariant partition function:
\ba
&&\hskip-10mm Z(\tau,\bar{\tau}) = \sum_{r,\bar{r}=1}^{m-1}
\sum_{s,\bar{s}=1}^{m+3}\,Z_0(\tau,\bar{\tau}) (N^{(m-2)}_{r,\bar{r}} 
N^{(m+2)}_{s,\bar{s}}
+N^{(m-2)}_{r,m-\bar{r}} N^{(m+2)}_{s,m+4-\bar{s}})\,
F^{(m)}_{rs}(\tau)\overline{F^{(m)}}_{\bar{r}\bar{s}}(\bar{\tau})~,
\nonumber \\
&&\hskip40mm s+r\equiv \bar{s}+\bar{r} \equiv 0~(\mbox{mod}~2).
\label{partG2}\ea
Here we may use any coefficient set of modular invariants 
$N^{(m-2)}_{r,\bar{r}}$, $N^{(m+2)}_{s,\bar{s}}$ 
of $SU(2)_{m-2}$, $SU(2)_{m+2}$ theories. $Z_0$ denotes the trivial part of 
the partition function which does not enter into the GSO projection
\ba
&&\hskip-5mm Z_0={1 \over |\prod_{n=1}(1-q^n)|^4}
\int_{-\infty}^{+\infty} dp\, dp_L \exp\left(-4\pi \tau_2\left({1 \over 2}
p^2+{1 \over 2}p_L^2+{1 \over 8}Q^2-{(c_L+1)\over 24}\right)\right)
 \nonumber \\
&&={1 \over \tau_2 |\eta(\tau)|^4} ,
\ea
where $c_L$ denotes the Liouville central charge $c_L=1+3Q^2$ and $\tau_2=
\mbox{Im} \,\tau$. $p_L\, (p)$ is the Liouville (Minkowski) momentum.
As is well-known, 
Liouville spectrum has a gap $h(L)\ge Q^2/8$.

We note that there exists a unique operator 
(in each chiral sector) which generates 
the analogue of the spectral flow in ${\cal N}=2$
theories between NS and Ramond sectors and is identified as the
space-time SUSY
operator in the above partition function. In fact the 
operator $\Psi=\phi^{(m)}_{1,2}\phi^{(m+2)}_{2,1}$ contained in R sector has a 
dimension $h^{(m)}_{1,2}+h^{(m+2)}_{2,1}=3/8$ for any values of $m$. When it is
properly dressed by the superconformal 
ghost and spin fields, it gives a current of conformal
dimension $1$
\be
J_{L,R}=e^{-\phi_{gh}/2}S_{\al}\Psi_{L,R} ,
\ee
(spin field 
$S_{\al}$ contains the contribution from the fermion of ${\cal N}=1$ Liouville
theory and has dimension $2/8$). 

Dimension of the fields $\phi_{r,p}^{(m)}\phi_{p,s}^{(m+2)}$ which appear in
the block $F^{(m)}_{r,s}$ is in general given by
\be
h^{r,s}_{p}={1 \over 4}\left(p-{r+s\over 2}\right)^2
+{((m+4)r-ms)^2-16\over 16m(m+4)}
+{\epsilon\over 8} ,
\ee
where $\epsilon=0$ for $p+r,p+s\equiv 0$ (mod 2) in NS sector and
$\epsilon=1$ for $p+r,p+s\equiv 1$ (mod 2) in R sector. Thus in the
``graviton orbit" $r=s=1$, 
\be
h^{1,1}_{p}={1\over 4}(p-1)^2+\frac{\epsilon}{8}, \hskip4mm p\le m+1 .
\ee
Hence fields in the NS sector of graviton orbit all possess integer conformal 
dimensions.
This suggests the existence of an extension of the chiral algebra
to some algebra involving higher spin fields
in our construction.
In fact $h^{r,s}_{p}$ and $h^{r,s}_{p\pm 2}$ differ by integers for any 
$p,r,s$ and it seems
quite likely that the sum over the product of minimal characters
$\sum_p \chi^{(m)*}_{r,p}\chi^{(m+2)*}_{p,s}$ provides a character
of an irreducible representation $(r,s)$ of the extended algebra.
Such an extended algebra for manifolds with exceptional holonomy was first
introduced by Shatashvili and Vafa \cite{SV} 
and further studied in refs.\cite{OF,Gep}.
We also note the pairing of NS and Ramond states 
\be
h^{r,s}_{p+1}-h^{r,s}_{p}={3\over 8}+{1\over 2}+
\mbox{integer}, \hskip5mm p=\mbox{odd}.
\ee 
The dimension $3/8$ is compensated by the spectral flow operator $\Psi$
and $1/2$ is consistent with  the GSO condition for NS sector
incorporated in the conformal blocks (\ref{cb G2}). 
If we recall the OPE of the minimal model
\be
\phi_{r,p}\phi_{1,2}\approx \phi_{r,p\pm 1},
\ee
we note that the operator $\Psi$ in fact generates a spectral flow.
We identify the state $r=s=1,p=3$ in the graviton orbit
as the associative 3-form $\Phi$ of the $G_2$ holonomy manifold since
it has dimension $3/2$ (contribution from the
fermions is added) and acts like the square of the spectral flow operator.

We also note that due to the presence of the gap in Liouville spectrum,
the dimension of the Ramond ground state satisfies an inequality
$h(m)+h(m+2)+h(L)\ge 1/24(c(m)+c(m+2)+c(L))=3/8$. 
Together with the contribution from 
$\theta_2$ it adds up to $1/2$ which is the value dictated by
space-time SUSY. Thus we believe that the partition function (\ref{partG2}) 
satisfies all the necessary
conditions for the string theory compactified on non-compact $G_2$ holonomy
manifolds. 

Except the special case of $m=3$ (\ref{cba G2}) does not appear
to contain tricritical Ising model. It turns out, however, due to some
theta function identities we can recast (\ref{cba G2}) into a form
which explicitly contain tricritical Ising model together with some
${\cal N}=0$ rational conformal models.\footnote
{We are grateful to C.Vafa for insisting on the existence of the tricritical
Ising model in the case of non-compact manifolds of $G_2$ holonomy. This 
prompted us to look for a formula involving tricritical Ising model and
correct an error in the original version of the manuscript.}
Explicit expressions are presented in 
section 4.

\section{Manifold with Spin(7) Holonomy}

~

\vskip-0.5cm

Next we would like to discuss a modular invariant for a non-compact
8-dimensional manifold with Spin(7) holonomy. Now we start from the following
identities
\ba
&&f_1(\tau)\equiv\frac{1}{2}\left\{\left(\frac{\th_3}{\eta}\right)^2
-\left(\frac{\th_4}{\eta}\right)^2\right\} \chi^{(1)}_{0}(\tau)
-\frac{1}{2}\left(\frac{\th_2}{\eta}\right)^2 \chi^{(1)}_{1}(\tau)=0~,
\label{identity spin7a}\\
&&f_2(\tau)\equiv\frac{1}{2}\left\{\left(\frac{\th_3}{\eta}\right)^2
+\left(\frac{\th_4}{\eta}\right)^2\right\} \chi^{(1)}_{1}(\tau)
-\frac{1}{2}\left(\frac{\th_2}{\eta}\right)^2 \chi^{(1)}_{0}
(\tau)=0~.
\label{identity spin7b}
\ea
Here $\chi^{(1)}_{\ell}$ is the character of 
level 1 $SU(2)$ WZW model with spin $\ell/2$ representation. 
(\ref{identity spin7a}),(\ref{identity spin7b}) are related to the
Jacobi's identity as
\be
{1 \over 2}\left(\left({\theta_3\over \eta}\right)^{\,4}
-\left({\theta_4\over \eta}\right)^{\,4}-\left({\theta_2
\over \eta}\right)^{\,4}\right)=
\chi^{(1)}_0f_1(\tau)+\chi^{(1)}_1f_2(\tau)
\label{identity spin7e}\ee
and were used previously in \cite{BG,Miz,ES}. 

In the present context we first decompose $\theta_i^2$ as ${\theta_i}^{1/2}
\times
\theta_i^{3/2}$ and replace $\theta_i^{3/2}$ by the level 2 $SU(2)$ characters
(\ref{branching N1-1}),(\ref{branching N1-2}),(\ref{branching N1-3}).
We then find the product of level 1 and 2 $SU(2)$ characters which 
gives rise to the tricritical Ising model under the standard 
coset construction of ${\cal N}=0$ minimal series
\be
{\cal M}^{\cN=0}_m: \hskip3mm {SU(2)_k\times SU(2)_1\over SU(2)_{k+1}},
\hskip3mm m=k+2.
\ee
Tricritical Ising model is identified as ${\cal M}^{\cN=0}_{m=4}$ and has
conformal blocks with dimensions
\ba
&&[h=0],\hskip3mm [h={1\over 10}], \hskip3mm [h={3\over 5}], 
\hskip3mm [h={3\over 2}] \\
&&[h={3\over 80}],\hskip3mm [h={7\over 16}].
\ea
It is known that tricritical Ising model possesses an ${\cal N}=1$ SUSY and is
also identified as the first one 
${\cal M}^{\cN=1}_{m=3}$ in the minimal ${\cal N}=1$ series. 
In ${\cal N}=1$ terms 
conformal blocks are organized as
\ba
\mbox{NS sector}:   \hskip4mm [h=0], \hskip3mm [h={1\over 10}],\\
\mbox{R sector}:   \hskip4mm [h={3 \over 80}], \hskip3mm [h={7\over 16}].
\ea

On the other hand the square roots of Jacobi theta functions may be identified
as characters of the Ising model
\ba
&&\chi^{\msc{Is}}_{\msc{h=0}}={1 \over 2}\left(\sqrt{{\theta_3 \over \eta}}
+\sqrt{{\theta_4 \over \eta}}\right)\\
&&\chi^{\msc{Is}}_{\msc{h=1/2}}={1 \over 2}\left(\sqrt{{\theta_3 \over \eta}}
-\sqrt{{\theta_4 \over \eta}}\right)\\
&&\chi^{\msc{Is}}_{\msc{h=1/16}}={1 \over \sqrt{2}}\sqrt{{\theta_2 \over \eta}}
\ea

Let us now introduce conformal blocks for the candidate partition function
of string compactified on a non-compact manifold of Spin(7) holonomy
\ba
&&F_1(\tau)\equiv\chi^{\msc{Is}}_{\msc{h=0}}\chi^{\msc{tri}}_{\msc{h=3/2}}+
\chi^{\msc{Is}}_{\msc{h=1/2}}\chi^{\msc{tri}}_{\msc{h=0}}-
\chi^{\msc{Is}}_{\msc{h=1/16}}\chi^{\msc{tri}}_{\msc{h=7/16}},\\
&&F_2(\tau)\equiv\chi^{\msc{Is}}_{\msc{h=0}}\chi^{\msc{tri}}_{\msc{h=1/10}}+
\chi^{\msc{Is}}_{\msc{h=1/2}}\chi^{\msc{tri}}_{\msc{h=3/5}}-
\chi^{\msc{Is}}_{\msc{h=1/16}}\chi^{\msc{tri}}_{\msc{h=3/80}}.
\ea
Identities (\ref{identity spin7a}),
(\ref{identity spin7b}) are then expressed as
\ba
&&f_1(\tau)=F_1(\tau)\chi^{(3)}_{0}(\tau)+F_2(\tau)\chi^{(3)}_{2}(\tau)=0,\label{identity spin7c}\\
&&f_2(\tau)=F_1(\tau)\chi^{(3)}_{3}(\tau)+F_2(\tau)\chi^{(3)}_{1}(\tau)=0.
\label{identity spin7d}\ea
Thus the branching functions $F_i\, (i=1,2)$ in fact vanish 
\be
F_i=0, \hskip4mm i=1,2
\label{spin7id}
\ee
consistently with the space-time SUSY. We have checked (\ref{spin7id})
by Maple. By construction blocks
$F_i$ have good modular properties. It is easy to derive the 
S-transformation of the conformal blocks
\be
\left(
\begin{array}{l}
 F_1\\
 F_2
\end{array}
\right) (-\frac{1}{\tau}) =
\frac{2}{\sqrt{5}}
\left(
\begin{array}{ll}
 \sin\left(\frac{\pi}{5}\right)&\sin\left(\frac{2\pi}{5}\right) \\
 \sin\left(\frac{2\pi}{5}\right)&-\sin\left(\frac{\pi}{5}\right)
\end{array}
\right) 
\left(
\begin{array}{l}
 F_1\\
 F_2
\end{array}
\right) (\tau) 
\ee

In the present case of 8-dimensional manifold 
there are no transverse degrees of freedom of Minkowski space. 
We interpret the Ising model sector as
arising from the Liouville fermion $\psi$. We note that 
there exists a space-time SUSY operator in the partition function:
a primary field $h=7/16$ in Ramond sector together with the superconformal
ghost and spin field gives a current $h=1$ (spin field contains a 
contribution from the Liouville fermion and has a dimension $3/16$).

Taking account of the contribution of the 
Liouville field the partition function
is then given by
\be
Z={1 \over \tau_2^{1/2}|\eta(\tau)|^2}\left(|F_1(\tau)|^2+|F_2(\tau)|^2
\right)
\ee

\section{Discussions}

In this article we have constructed candidate partition functions of
string theory compactified on non-compact manifolds with exceptional
holonomy. We have combined Liouville and ${\cal N}=1$ minimal models
so that the states in NS and Ramond sectors cancel 
and the theory possesses space-time supersymmetry. 
It seems quite likely that the manifolds of $G_2$ holonomy described by
our construction are the ALE spaces of $A_n$ type fibered over $S^3$.
The existence of $SU(2)$ current algebra originating from the
$SU(2)$ holonomy of ALE spaces is crucial in our construction.
We have found that pairs of ${\cal N}=1$ minimal models enter into
the partition function which contain spectral flow and SUSY
generators. Our construction is relatively easy since one
can freely adjust the central charge by Liouville field.

Let us now check that our amplitudes $F_{rs}^{(m)}$ (\ref{cba G2})
in fact contain tricritical Ising model as discussed in \cite{SV}. 
First we combine relations
(\ref{identity spin7e}),(\ref{identity spin7c}),(\ref{identity spin7d})
and obtain
\ba
&&\hskip-20mm
\chi^{(m-2)}_{r-1}{1 \over \eta^4}(\theta_3^4-\theta_4^4-\theta_2^4)
=2\chi^{(m-2)}_{r-1}(\chi^{(1)}_0\chi^{(3)}_0+\chi^{(1)}_1\chi^{(3)}_3)F_1
\nonumber \\
&&\hskip50mm 
+2\chi^{(m-2)}_{r-1}(\chi^{(1)}_0\chi^{(3)}_2+\chi^{(1)}_1\chi^{(3)}_1)F_2.
\ea
We then introduce the branching functions $\chi^{(M;L)}_{(r,s;t)}(\tau)$
for the cosets $SU(2)_{M}\times SU(2)_L
/SU(2)_{M+L}$, 
\ba
&&\chi^{(M)}_{r-1}(\tau)\chi^{(L)}_{t}(\tau)
=\sum_{\stackrel{s=1}{|r-s-t|\equiv 0 \,(\msc{mod 2})}}^{M+L+1}
\chi^{(M;L)}_{(r,s;t)}(\tau)\chi^{(M+L)}_{s-1}(\tau),\\
&&\hskip30mm (1\le r\le M+1,\, 1\le s \le M+L+1,\, 0\le t\le L), \nonumber\\
&&\hskip30mm \chi^{(M;L)}_{(r,s;t)}=\chi^{(M;L)}_{(M-r+2,M+L-s+2;L-t)},
\ea
and obtain
\ba
&&\hskip-10mm
\chi^{(m-2)}_{r-1}{1 \over \eta^4}(\theta_3^4-\theta_4^4-\theta_2^4)
=2\sum_{s=1}^{m+3}\chi^{(m+2)}_{s-1}(\tau)
\left[F_1(\tau)\left\{\sum_p\chi^{(m-2;1)}_{(r,p;0)}
\chi^{(m-1;3)}_{(p,s;0)}+\sum_{p'}\chi^{(m-2;1)}_{(m-r,p';0)}
\chi^{(m-1;3)}_{(p',m+4-s;0)}\right\}(\tau)\right.\nonumber \\
&&\hskip30mm \left.+F_2(\tau)\left\{\sum_p\chi^{(m-2;1)}_{(r,p;0)}
\chi^{(m-1;3)}_{(p,s;2)}+\sum_{p'}\chi^{(m-2;1)}_{(m-r,p';0)}
\chi^{(m-1;3)}_{(p',m+4-s;2)}\right\}(\tau)\right].
\ea
By comparing with (\ref{identity G2}) we find
\ba
&&F^{(m)}_{rs}(\tau)=F_1(\tau)\left\{\sum_p\chi^{(m-2;1)}_{(r,p;0)}
\chi^{(m-1;3)}_{(p,s;0)}+\sum_{p'}\chi^{(m-2;1)}_{(m-r,p';0)}
\chi^{(m-1;3)}_{(p',m+4-s;0)}\right\}(\tau)\nonumber \\
&&\hskip30mm +F_2(\tau)\left\{\sum_p\chi^{(m-2;1)}_{(r,p;0)}
\chi^{(m-1;3)}_{(p,s;2)}+\sum_{p'}\chi^{(m-2;1)}_{(m-r,p';0)}
\chi^{(m-1;3)}_{(p',m+4-s;2)}\right\}(\tau).
\label{tci rep}\ea
Thus our conformal blocks are now expressed
in terms of tricritical Ising model together with some 
combinations of ${\cal N}=0$ rational conformal field theories.
This description seems to fit exactly with the discussion given by
Shatashvili and Vafa \cite{SV}.
In the right-hand-side of 
(\ref{tci rep}) ${\cal N}=1$ world-sheet SUSY is not
manifest and also the Liouville fermion is not explicit.
In our original representation (\ref{cba G2}), on the other hand,
world-sheet SUSY and the Liouville fermion are manifest but
the tricritical Ising model is difficult to identify.

In the case of compact manifolds
one can no longer use the Liouville field and can not
freely adjust the central charge of the theory: the
system is much more rigid. 
Construction of modular invariants for compact manifolds
is a challenging problem which we would like
to address in future communications.

\vskip1.5cm

\section*{Acknowledgement}

~
\vskip-0.5cm

We would like to dedicate this article to the memory of our friend and 
colleague, Dr. Sung-Kil Yang of Tsukuba University
who recently passed away after an illness of 1 and 1/2 years.
He will be greatly missed by all who knew him.

We thank the organizers of Summer Institute 2001 at Lake Kawaguchi
for its stimulating atmosphere where this work was completed.
Research of T.E. and Y.S. are supported in part by the fund,
Special Priority Area No.707 ``Supersymmetry and Unified Theory of
Elementary Particles", Japan Ministry of Education and Science.

\bigskip

\bigskip

\newpage

\section*{Appendix: ~ ${\cal N}=1$ Unitary Minimal Models }

\setcounter{equation}{0}
\def\theequation{A.\arabic{equation}}

~

\vskip-0.5cm

We summarize basic data on ${\cal N}=1$ Virasoro minimal models.

\begin{enumerate}
\item Central charge and conformal dimensions:
 \begin{eqnarray}
 &&c=\frac{3}{2}-\frac{12}{m(m+2)},~~~ (m=3,4,5,\ldots)
 \label{center N1} \\
 &&h^{(m)}_{r,s} = \frac{((m+2)r-ms)^2-4}{8m(m+2)}
 +\frac{1-(-1)^{r-s}}{32}   , \\
 &&\hskip4mm \equiv\frac{((m+2)r-ms)^2-4}{8m(m+2)}+\frac{\ep}{16} \nonumber \\
 && \nonumber \\
 &&\ep =
 \left\{
 \begin{array}{ll}
 0& r+s\equiv 0 ~(\mod ~2) ~:\hskip3mm \mbox{NS-sector} \\
 1& r+s\equiv 1 ~(\mod ~2) ~:\hskip3mm \mbox{R-sector}
 \end{array}
 \right.  \\
 && 
 ~~~1\leq r \leq m-1,~ 1\leq s \leq m+1  \nonumber \\
 && \nonumber \\
 &&h^{(m)}_{m-r,m+2-s}=h^{(m)}_{r,s} ~.
 \end{eqnarray}
\item
 Character formulas \cite{GKO}:
 \begin{eqnarray}
 \chi^{(m)\, NS}_{r,s}(\tau) &=&\frac{1}{\eta}\sqrt{\frac{\th_3}{\eta}}\, 
 K^{(m)}_{r,s}(\tau) \equiv q^{-\frac{1}{16}}\prod_{n=1}^{\infty}
 \frac{1+q^{n-\frac{1}{2}}}{1-q^n}\,K^{(m)}_{r,s}(\tau),
 \label{character NS-1}\\
 K^{(m)}_{r,s}(\tau)&\df& \Th{(m+2)r-ms}{2m(m+2)}(\tau)
 +\Th{(m+2)r-ms+2m(m+2)}{2m(m+2)}(\tau) \nonumber\\
 && -\Th{(m+2)r+ms}{2m(m+2)}(\tau)
 -\Th{(m+2)r+ms+2m(m+2)}{2m(m+2)}(\tau) \nonumber\\
 &=& \Th{(m+2)r-ms}{m(m+2)}(\tau/2)-\Th{(m+2)r+ms}{m(m+2)}(\tau/2)~.\\
 && \nonumber \\
 \tchi^{(m)\, NS}_{r,s}(\tau) &=&\frac{1}{\eta}\sqrt{\frac{\th_4}{\eta}}\, 
 \tK^{(m)}_{r,s}(\tau)\equiv q^{-\frac{1}{16}}\prod_{n=1}^{\infty}
 \frac{1-q^{n-\frac{1}{2}}}{1-q^n}\,\tK^{(m)}_{r,s}(\tau),
 \label{character NS-2}\\
 \tK^{(m)}_{r,s}(\tau)&\df&(-1)^{\frac{r-s}{2}} \Th{(m+2)r-ms}{2m(m+2)}(\tau)
 \nonumber\\
 &&+(-1)^{\frac{r-s}{2}+m} \Th{(m+2)r-ms+2m(m+2)}{2m(m+2)}(\tau) \nonumber\\
 && -(-1)^{\frac{r+s}{2}}\Th{(m+2)r+ms}{2m(m+2)}(\tau) \nonumber\\
 && -(-1)^{\frac{r+s}{2}+m} \Th{(m+2)r+ms+2m(m+2)}{2m(m+2)}(\tau) ~.\\
 && \nonumber \\
 \chi^{(m)\,R}_{r,s}(\tau) &=& \frac{1}{\sqrt{2}\eta}
 \sqrt{\frac{\th_2}{\eta}}\, K^{(m)}_{r,s}(\tau)
 \equiv \prod_{n=1}^{\infty}\frac{1+q^n}{1-q^n}\, K^{(m)}_{r,s}(\tau)~
 \label{character R}
 \end{eqnarray}
\end{enumerate}


\end{document}